\documentclass[aps,reprint,superscriptaddress,prb]{revtex4-1}
\usepackage{amssymb}
\usepackage{amsmath}
\usepackage{graphicx}
\usepackage{mathbbol}

\newcommand{\rmi}{\ensuremath{\mathrm{i}}}

\newcommand{\rmd}{d}

\usepackage[english]{babel}

\newcommand{\energy}{E}

\begin{document}
\title{The effect of dynamical Bloch oscillations on
  optical-field-induced current in a wide-gap dielectric}

\author{P\'{e}ter F\"{o}ldi}
\email{foldi@physx.u-szeged.hu}
\affiliation{Department of Theoretical Physics, University of Szeged,
  Tisza Lajos k\"{o}r%
  \'{u}t 84-86, H-6720 Szeged, Hungary}

\author{Mih\'aly G. Benedict}
\affiliation{Department of Theoretical Physics, University of Szeged,
  Tisza Lajos k\"{o}r%
  \'{u}t 84-86, H-6720 Szeged, Hungary}

\author{Vladislav S. Yakovlev}
\email{vladislav.yakovlev@lmu.de}
\affiliation{Ludwig-Maximilians-Universit\"at, Am Coulombwall 1, 85748 Garching, Germany}
\affiliation{Max-Planck-Institut f\"ur Quantenoptik, Hans-Kopfermann-Stra{\ss}e 1, 85748 Garching, Germany}

\begin{abstract}
  We consider the motion of charge carriers in a bulk wide-gap
  dielectric interacting with a few-cycle laser pulse. A semiclassical
  model based on Bloch equations is applied to describe the emerging
  time dependent macroscopic currents for laser intensities
  close to the damage threshold. At such laser intensities,
  electrons can reach edges of the first Brillouin zone even for
  electron--phonon scattering rates as high as those known for
  $\mbox{SiO}_2$. We find that, whenever this happens, Bragg-like
  reflections of electron waves, also known as Bloch oscillations,
  affect the dependence of the charge displaced by the laser pulse on
  its carrier--envelope phase.
\end{abstract}

\pacs{72.40.+w, 78.20.Bh}
\maketitle

\section{Introduction}
The motion of conduction-band electrons in a crystalline solid is
usually considered to be similar to that in free space, apart from
scattering processes. The situation is radically different in the case
where an external electric field is so strong that, in spite of
scattering, an electron can acquire such a high crystal momentum that
it reaches an edge of the first Brillouin zone. While the kinetic
energy of a free electron exposed to a constant external field would
indefinitely increase, an electron in a crystal first slows down until
it reaches the top of the energy band, and then it moves in the
opposite direction towards the bottom of the band. In the
semiclassical picture neglecting scattering, the electron would move
periodically back and forth between its initial and final positions.
This phenomenon is known as Bloch oscillations~\cite{Bloch1929}, and
it leads to Wannier--Stark localization~\cite{Wannier1960}. In the
reduced band scheme, an electron reaching an edge of the first
Brillouin zone continues its motion from the opposite side of the
zone. In the real space, this corresponds to a Bragg-like reflection
of an electron wave~\cite{Houston1940}. While Bloch oscillations and
Wannier--Stark localization are usually considered in the case of a
constant external field, essentially the same physical phenomena take
place if the external field is time dependent. In this paper, we use
the term 'dynamical Bloch oscillations' to describe phenomena
that occur whenever an electron wave is reflected at an edge of the
Brillouin zone.

Until recently, Bloch oscillations were thought to be impossible to
observe in bulk solids because of scattering. The period of Bloch
oscillations in a constant field $F$ is given by $T_\mathrm{B}=h (e F
a)^{-1}$, where $h$ is the Planck constant and $a$ is a lattice
period. To observe Bloch oscillations, $T_\mathrm{B}$ must be smaller
than characteristic scattering and dephasing times, which are usually
on the order of $T_\mathrm{s} \sim 10^{-13}$~s. This implies that the
external field must be stronger than $F \gtrsim h(e a
T_\mathrm{s})^{-1} \sim 10^8\ \mbox{V}\,\mbox{m}^{-1}$. Such a strong
constant field would destroy even wide-gap dielectrics. Therefore,
over the last few decades, Bloch oscillations were predominantly
studied in artificial periodic structures, such as semiconductor
superlattices~\cite{Mendez1988,Feldmann1992}, a notable exception
being the observation of partial Bloch oscillations in $n$-doped GaAs
interacting with intense terahertz pulses~\cite{Kuehn2010}. A closely
related phenomenon, optical Bloch oscillations, was observed in
periodic waveguide structures~\cite{Pertsch1999,Morandotti1999},
periodic dielectric systems~\cite{Sapienza2003} and optical lattices
fabricated from porous silicon~\cite{Agarwal2004}. Also, Bloch
oscillations were also studied for ultracold atoms in optical
superlattices~\cite{Dahan1996,Holthaus2000}.

The situation has recently changed as intense few-cycle pulses were
generated in the mid-infrared (MIR) spectral region~\cite{Gruetzmacher2002},
and the duration of the shortest near-infrared (NIR) pulses approached
one optical cycle~\cite{Goulielmakis2008}.
Ghimire \emph{et al.}~\cite{Ghimire_NP_2011,Ghimire_PRL_2011} observed that
anharmonicity in the motion of charge carriers created and driven by an intense MIR
pulse in ZnO resulted in the generation of high-order harmonics~\cite{Golde2008}
and a red-shift of absorption edge~\cite{Ghimire_PRL_2011}. In their
parameter regime, the field was intense enough to drive conduction
electrons beyond the first Brillouin zone, so that Bloch oscillations
were suggested to be responsible for the observed effects.

Very recently, Schiffrin \emph{et al.}~\cite{Schiffrin2013} found that a
4 fs NIR pulse with a peak electric field of $20\
\mbox{GV}\,\mbox{m}^{-1}$ can induce measurable currents in a
$\mbox{SiO}_2$ sample. Furthermore, it has been found that these
currents can be steered by controlling the carrier--envelope phase
(CEP)~\cite{Baltuska2003} of laser pulses. These findings were
interpreted in terms of Wannier--Stark states~\cite{Apalkov2012}.

While it remains debatable whether Bloch oscillations played a major
role in these particular measurements, there is no doubt that intense
few-cycle pulses enable experiments in the parameter regime where
electrons in a bulk solid are accelerated beyond the first Brillouin
zone. The purpose of this paper is to study some basic effects related
to this parameter regime. In particular, we consider the role of
electron scattering and dephasing.

Electron--phonon scattering rates are known to be particularly high
for $\mbox{SiO}_2$ due to a strong coupling between conduction
electrons and longitudinal optical (LO) phonons~\cite{Hughes1973,Fischetti1985}.
At the same time, we are not aware of any direct measurements of
scattering rates for moderately hot ($\energy_\mathrm{kin} \sim 1\
\mbox{eV}$) electrons, which we consider in this paper, and there are
still open questions related to the scattering of very hot conduction
electrons~\cite{Quere2000}.

\section{The system and the model}
\begin{figure}[htb]
\begin{center}
\includegraphics[width=0.45\textwidth]{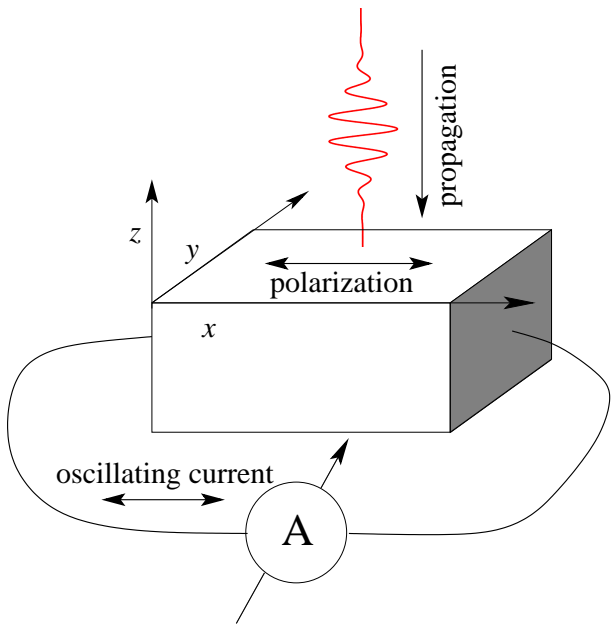}
\caption{A schematic view of the creation and driving of macroscopic
  currents with a laser pulse. Measurements~\cite{Schiffrin2013} yield
  the charge transferred by the pulse.}
\label{schemefig}
\end{center}
\end{figure}

We consider the following model (see Fig.~\ref{schemefig}): the $xy$
$(z=0)$ plane is the surface of a dielectric. Short pulses with a
stabilized CEP propagate along the $z$-axis and impinge on this
surface. In our calculations, we assume that the external electric
field within the sample is linearly polarized and given by
$\mathcal{E}_x(t)=\mathcal{E}_0\cos(\omega_0 t+\varphi_{\mathrm{CEP}})
\exp[-t^2/(2\tau^2)]$, $\mathcal{E}_y=\mathcal{E}_z=0$.

We adopt the two-band approximation and consider the electron motion
in two spatial dimensions. At each moment $t$, we describe electronic
excitations in the sample with the aid of quantum-mechanical density
matrices
\begin{equation}\label{rho}
  \rho(\mathbf{k},t)=\left( \begin {array}{cc} n_\mathrm{c}(\mathbf{k},t) &\mathcal{P}(\mathbf{k},t) \\
      \mathcal{P}^*(\mathbf{k},t) &n_\mathrm{v}(\mathbf{k},t) \end{array}  \right),
\end{equation}
where $n_\mathrm{c}$, $n_\mathrm{v}$ correspond to the conduction and
valence band populations, the off-diagonal element
$\mathcal{P}(\mathbf{k},t)$ represents the interband coherence and the
crystal momentum $\mathbf{k}$ has two components: $\mathbf{k}=(k_x, k_y)$. The
time dependence of the density matrix can be formally written as
\begin{equation}
\label{formaleq}
\frac{\partial {}}{\partial t}\rho(\mathbf{k},t)=
\left(\frac{\partial \rho}{\partial t}\right)_{\mathrm{exc}}
+\left(\frac{\partial \rho}{\partial t}\right)_{\mathrm{force}}
+ \left(\frac{\partial \rho}{\partial t}\right)_{\mathrm{scatt}}.
\end{equation}
The three terms on the right-hand side of this equation describe the
effects that we take into account: photoexcitation, the acceleration
of charge carriers by the external field and electron scattering,
which acts as a 'friction force'. In the following, we use well
established models to account for each of these phenomena; however,
our model is rather phenomenological as we did not systematically
derive it from first principles. The last two terms on the right-hand
side of Eq.~\eqref{formaleq} are obtained from the standard single-band
Boltzmann equation~\cite{Kittel2005}. The role of the laser field is
twofold here: besides driving charge carriers in the
conduction and valence bands
$\left[\left(\partial_t \rho\right)_{\mathrm{force}}\right]$,
it also drives interband transitions, i.e., populates the initially
empty conduction band
$\left[\left(\partial_t \rho\right)_{\mathrm{exc}}\right]$.

It is common to describe strong-field excitations using rates for
multiphoton or tunnelling transitions, but the applicability of this
approach is very questionable for extremely short laser pulses at
intensities where the Keldysh parameter is comparable to 1. Therefore,
we use a quantum-mechanical model for the term
$\left(\partial_t \rho\right)_{\mathrm{exc}}$
in Eq.~\eqref{formaleq}, describing photoexcitation with $\mathbf{k}$-resolved
optical Bloch equations~\cite{Casperson_1998,HaugKoch2004} in the
two-band approximation:
\begin{eqnarray}
\label{bloch}
\left(\frac{\partial \mathcal{P}(\mathbf{k})}{\partial t}\right)_{\mathrm{exc}}=&-&\frac{\rmi}{\hbar}\left [\energy_\mathrm{c}(\mathbf{k})-\energy_\mathrm{v}(\mathbf{k}) -\rmi \hbar\kappa\right]\mathcal{P}(\mathbf{k})  \nonumber\\
&-&\rmi[n_\mathrm{c}(\mathbf{k})-n_\mathrm{v}(\mathbf{k})]d_\mathrm{c v}(\mathbf{k})\mathcal{E}(t), \nonumber\\
\left(\frac{\partial n_\mathrm{c}(\mathbf{k})}{\partial t}\right)_{\mathrm{exc}}=&-&2\,\mathrm{Im}[d_\mathrm{c v}(\mathbf{k})\mathcal{E}(t)\mathcal{P}^*(\mathbf{k})], \nonumber\\
\left(\frac{\partial n_\mathrm{v}(\mathbf{k})}{\partial t}\right)_{\mathrm{exc}}=&-&\frac{\partial n_\mathrm{c}(\mathbf{k})}{\partial t}.
\end{eqnarray}
Here, $\energy_\mathrm{v}(\mathbf{k})$ and $\energy_\mathrm{c}(\mathbf{k})$
are the energies of the valence and conduction bands, respectively,
$d_\mathrm{c v}(\mathbf{k})$ is the $x$--component of the interband transition matrix
element, and the phenomenological rate $\kappa$ describes the decay of
the interband coherences. We neglect interband relaxation (population
decay) as it occurs on the picosecond to nanosecond time scale. For
simplicity, we estimate the dependence of the dipole matrix elements
on $\mathbf{k}$ as
$ d_\mathrm{c
  v}(\mathbf{k})=d_\mathrm{c v}(0)
\frac{\energy_\mathrm{c}(0)-\energy_\mathrm{v}(0)}
{\energy_\mathrm{c}(\mathbf{k})-\energy_\mathrm{v}(\mathbf{k})}$
(see~Ref.~\onlinecite{HaugKoch2004}). The actual
value of $d_\mathrm{c v}(0)$ has a minor qualitative effect on our
results, as long as saturation-related phenomena are negligible, i.e.\
the excited population is well below unity. In the following, we use
$d_\mathrm{c v}(0)=0.1$ atomic units. Note that Eqs.~\eqref{bloch}
make no use of the rotating wave approximation. This allows us to
investigate dynamics that unfold within a single optical oscillation
of the laser pulse (e.g.~Ref.~\onlinecite{Wegener_2004} contains a detailed
discussion of related phenomena).

The external electric field not only causes transitions between the
bands, but it also accelerates and decelerates charge carriers. These
field-driven dynamics are accounted for by the second term in
Eq.~\eqref{formaleq}. We neglect off-diagonal terms in
$\left(\partial_t \rho\right)_{\mathrm{force}}$
and evaluate the diagonal ones as
\begin{equation}
  \left(\frac{\partial n_\mathrm{v,c}(\mathbf{k})}{\partial
      t}\right)_{\mathrm{force}}=-\frac{e}{\hbar}\mathbf{\mathcal{E}}(t)
  \nabla_{\mathbf{k}} n_\mathrm{v,c}(\mathbf{k}).
\label{fieldeq}
\end{equation}
Note that $\mathbf{\mathcal{E}}(t)$ is the electric field \emph{in the
  medium}, so that it is assumed to include both the screening field
due to the collective electron response~\cite{Otobe2008} and the field
due to the polarization of the sample.

The third term in Eq.~\eqref{formaleq} accounts for the loss of intraband
coherence due to scattering, where longitudinal optical (LO) phonons
are considered to play the major role. The electron--phonon interaction
is described by the Fr\"{o}hlich Hamiltonian, and standard
methods~\cite{HaugKoch2004} lead to the following dynamical equations:
\begin{widetext}
\begin{multline}
  \label{scatteq}
  \left(\frac{\partial n_\mathrm{c}(\mathbf{k})}{\partial t}\right)_{\mathrm{scatt}}=\gamma_0\sum_{\mathbf{q}} \delta \bigl(E(\mathbf{k+q})-\energy_\mathrm{c}(\mathbf{k})-\hbar \omega_\mathrm{LO}\bigr)
  \frac{1}{q^2} \biggl\{-N_{\mathbf{q}}n_\mathrm{c}(\mathbf{k})\left[1-n_\mathrm{c}(\mathbf{k+q}) \right] +
  (N_{\mathbf{q}}+1)n_\mathrm{c}(\mathbf{k+q})\left[1-n_\mathrm{c}(\mathbf{k})\right]\biggr\} + \\
  \gamma_0\sum_{\mathbf{q}} \delta \bigl(E(\mathbf{k-q})-\energy_\mathrm{c}(\mathbf{k})+\hbar \omega_\mathrm{LO}\bigr)
  \frac{1}{q^2}\biggl\{-(N_{\mathbf{q}}+1)n_\mathrm{c}(\mathbf{k})\left[1-n_\mathrm{c}(\mathbf{k-q}) \right]+
  N_{\mathbf{q}}n_\mathrm{c}(\mathbf{k-q})\left[1-n_\mathrm{c}(\mathbf{k}) \right] \biggr\},
\end{multline}
\end{widetext}
where $\hbar\omega_\mathrm{LO}$ denotes the energy of an LO phonon,
which is assumed to be independent of the reciprocal-space vector
$\mathbf{q}$. The related phonon density is denoted by
$N_{\mathbf{q}}$, and $\gamma_0$ is an electron--phonon coupling
constant. For the sake of simplicity, we use a tight-binding-type
dispersion relation, i.e.\ the energies
$\energy_\mathrm{c,v}(\mathbf{k})-\energy_\mathrm{c,v}(0)$ are
proportional to $2-\cos(k_xa)-\cos(k_ya)$, where $a$ is a lattice
constant. At room temperature, $N_{\mathbf{q}}$ is practically zero,
so phonon emission processes (accompanied by electron scattering
events with a loss of electron energy) dominate the scattering
dynamics. Note that the scattering process described above
qualitatively depends on the number of dimensions. By using a
one-dimensional model, we would strongly underestimate momentum
relaxation, as the probability of back-scattering $(q \approx 2 k)$ is
negligibly small for electrons with a kinetic energy larger than a
fraction of electronvolt (eV). If there is more than one spatial
dimension, electron deflection upon scattering results in a faster
decrease of the net momentum. We chose to use two spatial dimensions
in our calculations as a compromise between building a possibly
realistic model for electron--phonon scattering and keeping
computational time at an acceptable level.

The model described above allows us to calculate the rate of
scattering on LO phonons $\gamma(\mathbf{k})$, which is proportional to
$\gamma_0$ appearing in Eq.~\eqref{scatteq}. This scattering rate
determines how fast the crystal momentum of an electron wave packet
with a well-defined $\mathbf{k}$ decreases as a consequence of scattering
events. Our numerical calculations show, in accordance with the
analytical results presented e.g.\ in~Ref.~\onlinecite{Fischetti1985}, that the
rate $\gamma(\mathbf{k})$ has a pronounced minimum at $\mathbf{k}=0$, being
nearly constant ($\gamma(\mathbf{k}) \approx \gamma$) for kinetic energies
in the range between 0.5 eV and 2 eV. In the following, we use
$\gamma$ as a label to quantify the strength of electron--phonon
interaction in different simulations, although we used
$\mathbf{k}$-dependent scattering rates in our calculations. It must also be
mentioned that the overall momentum of electrons distributed over a
large part of the Brillouin zone may decrease at a rate that is much
smaller than $\gamma$ because, for a broad electron distribution,
scattering events that increase the net momentum are approximately as
probable as those that decrease it.

For our simulations, we used material parameters that correspond to
SiO${}_2$: a band gap of 9 eV, lattice period $a=0.5\ \mbox{nm}$, two
LO phonon modes with energies $\hbar\omega_\mathrm{LO}=0.153$ and
0.063 eV, and a combined scattering rate equal to $\gamma=0.3\
\mbox{fs}^{-1}$. The laser pulse parameters correspond to the
experiment described in~Ref.~\onlinecite{Schiffrin2013}: $\omega_0=2.51\
\mbox{fs}^{-1}$ and $\tau=2.3\ \mbox{fs}$ ($\mbox{FWHM}=3.8\ \mbox{fs}$).
Note that in this case the band gap is more than five times larger
than $\hbar\omega_0$; thus, the excitation is far from being resonant
or, in other words, it is a multiphoton process.
\begin{figure}[htb]
\begin{center}
\includegraphics[width=0.45\textwidth]{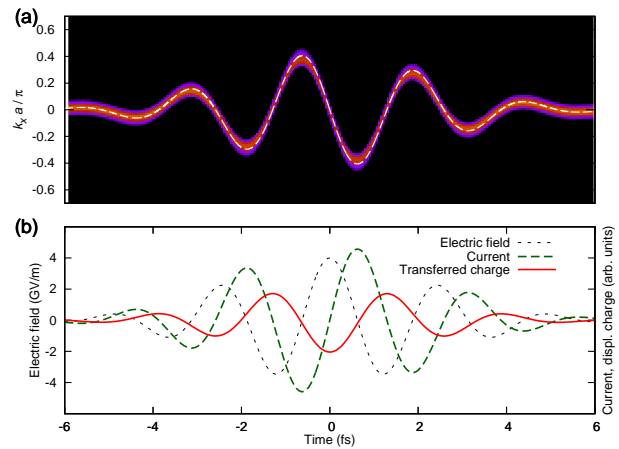}
\caption{Laser-driven motion of conduction-band electrons that
  initially have a Gaussian distribution:
  $n_\mathrm{c}(\mathbf{k},t_\mathrm{min})=\exp\left[-(k_x^2+k_y^2) a^2
    \Delta^{-2}\right]$ with $\Delta=0.03.$ This simulation neglects
  photoexcitation and electron--phonon scattering. (a) The
  distribution of conduction electrons $n_\mathrm{c}(\mathbf{k},t)$ in the
  $k_y=0$ plane. The dashed white line is the 'semiclassical
  trajectory' evaluated with the aid of the acceleration theorem
  \eqref{acceleration_theorem}. (b) The electric field
  $\mathcal{E}_x(t)$ that drives the wave packet, the current density
  $j_x(t)$ and the time dependent transferred charge $Q(t)$. The
  parameters are $\mathcal{E}_0=4\ \mbox{GV}\,\mbox{m}^{-1}$,
  $\varphi_{\mathrm{CEP}}=0$, $\tau=2.3\ \mbox{fs}$.}
\label{fieldonlyfig}
\end{center}
\end{figure}

\begin{figure}[htb]
\begin{center}
  \includegraphics[width=0.45\textwidth]{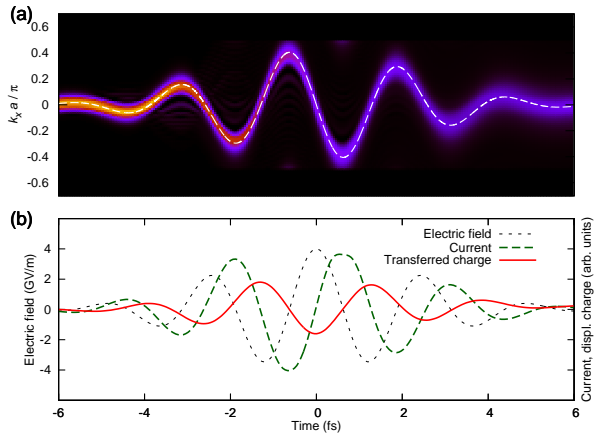}
  \caption{Time evolution of a wave packet initially centred at
    $\mathbf{k}=0$ in a simulation that neglects photoexcitation but takes
    electron--phonon scattering into account. $\gamma=0.3\
    \mbox{fs}^{-1}$; all other parameters are the same as in
    Fig.~\ref{fieldonlyfig}. Electron scattering along both $k_x$
    and $k_y$ is responsible for the gradual reduction of the
    reciprocal-space electron distribution $n_\mathrm{c}(\mathbf{k},t)$ towards
    the end of the laser pulse, but it has a relatively weak effect on
    $j_x(t)$.}
\label{fieldonlyfig2}
\end{center}
\end{figure}

\section{Results}
\subsection{Electron acceleration in the laser field}
Before we present simulations where a laser pulse creates and drives
charge carriers, let us consider the laser-driven motion of
initially free electrons neglecting interband transitions. Similar
simulations can be found in~Ref.~\onlinecite{Muecke2011}. As an example, we take
the initial distribution of conduction electrons as a Gaussian wave
packet centred at $\mathbf{k}=0$, neglect the terms
$\left(\partial_t \rho\right)_{\mathrm{exc}}$
and
$\left(\partial_t \rho\right)_{\mathrm{scatt}}$
in Eq.~\eqref{formaleq} and model the time evolution of the electron wave
packet by solving Eq.~\eqref{fieldeq}. We plot the distribution of
conduction electrons in false-colour diagrams, where colours vary from
black through red to yellow as the electron population
$n_\mathrm{c}(\mathbf{k},t)$ grows from zero to its maximal value. The top
panel of Fig.~\ref{fieldonlyfig} presents such a diagram in the
plane $k_y=0$. In this example, the distribution remains localized in
the reciprocal space, and it is dynamically shifted by the field of
the laser pulse. The reciprocal-space motion of the wave packet is
appropriately described by the 'acceleration theorem'~\cite{Kittel2005}:
\begin{equation}
  \label{acceleration_theorem}
  \frac{\partial \mathbf{k}}{\partial t} = -\frac{e}{\hbar} \mathbf{\mathcal{E}}(t).
\end{equation}
A solution of this equation with the initial condition
$k(t_\mathrm{min})=0$ is shown by the dashed white line in Fig.~\ref{fieldonlyfig}.
This solution can be regarded as a trajectory of a
'classical particle' in the reciprocal space. Let us note that as long
as neither scattering nor excitation is taken into account, there is a
simple scaling in the model: increasing both the carrier frequency
$\omega_0$ and the amplitude of the laser pulse $\mathcal{E}_0$ by a
certain factor is equivalent to choosing a new unit of time in
Eq.~\eqref{formaleq}, and it does not change the maximal crystal momentum
that a wave packet can reach in the reciprocal space.

Having evaluated the time evolution of the density matrix, we are able to
investigate measurable physical quantities.

The number of electrons excited per unit cell is given by
\begin{equation}
  \label{nc}
  \langle n_\mathrm{c} \rangle = \Omega^{-1} \int n_\mathrm{c}(\mathbf{k})\,\rmd^2 \mathbf{k}
\end{equation}
with
\begin{equation}
  \label{eq:Omega}
  \Omega = \int \rmd^2 \mathbf{k}.
\end{equation}
The electric current per unit cell $\mathbf{j}(t)$ is a sum of
contributions from conduction-band electrons
\begin{equation}
  \label{j1}
  \mathbf{j}_\mathrm{c} (t)=\Omega^{-1} \int_\mathrm{BZ} e
  \mathbf{v}_\mathrm{c}(\mathbf{k}) n_\mathrm{c}(\mathbf{k},t)\,\rmd^2\mathbf{k},
\end{equation}
valence-band holes
\begin{equation}
  \label{j2}
  \mathbf{j}_\mathrm{v} (t)=\Omega^{-1} \int_\mathrm{BZ} e
  \mathbf{v}_\mathrm{v}(\mathbf{k}) n_\mathrm{v}(\mathbf{k},t)\,\rmd^2\mathbf{k},
\end{equation}
and interband coherences
\begin{multline}
  \label{j12}
  \mathbf{j}_\mathrm{c v} (t)=2 \hbar^{-1} \Omega^{-1}\,\mathrm{Im} \Biggl\{\int_\mathrm{BZ}
    \mathcal{P}(\mathbf{k},t) \mathbf{d}_\mathrm{c v}(\mathbf{k})\times\\
    \left[\energy_v(\mathbf{k})-\energy_c(\mathbf{k})\right]\,\rmd^2\mathbf{k}\Biggr\},
\end{multline}
where the integrals are taken over the first Brillouin zone (in the
$k_x$ and $k_y$ directions, in accordance with our 2D model), and the
velocity distributions are given by
$\mathbf{v}_\mathrm{v,c}(\mathbf{k})=\hbar^{-1}\nabla_{\mathbf{k}}
\energy_\mathrm{v,c}(\mathbf{k})$. Equations~\eqref{j1}-\eqref{j12} result
from evaluating the expectation value of the current operator averaged
over a unit cell for a state described by the density matrix
\eqref{rho}. Integrating the current density with respect to time, we
obtain the charge (per unit cell) that flows through a surface
perpendicular to the $x$--axis: $Q(t)=\int_{-\infty}^t j_x(t')\,\rmd
t'=Q_\mathrm{c}(t)+Q_\mathrm{v}(t)+Q_\mathrm{c v}(t)$, where $j_x$
denotes the $x$--component of the total current density
$\mathbf{j}=\mathbf{j}_\mathrm{c}+\mathbf{j}_\mathrm{v}+\mathbf{j}_\mathrm{c v}.$ It
is the charge displaced by the laser pulse $Q=Q(\infty)$ that can be
measured in experiments~\cite{Schiffrin2013}. In the following, we
assume that $\mathbf{j}_\mathrm{c}$ gives the dominant contribution to
$Q$. Indeed, $\mathbf{j}_\mathrm{v}$ is negligible in comparison with
$\mathbf{j}_\mathrm{c}$ due to the low mobility of holes in the valence
band, while $\mathbf{j}_\mathrm{c v}$ mainly describes the polarization
response of valence-band electrons; according to our calculations,
$Q_\mathrm{c v}(t)$ is proportional to the applied field within a
relative error of no more than 5\% up to a field amplitude of
$\mathcal{E}_0=25\ \mbox{GV}\,\mbox{m}^{-1}$, $Q_\mathrm{c v}(\infty)$
being negligibly small in comparison to $Q_\mathrm{c}(\infty)$. Keeping
this in mind, we restrict our analysis to $\mathbf{j}_\mathrm{c}$ and
$Q_\mathrm{c}$, referring to them as 'current density' and
'transferred charge', respectively. As long as the laser field is
linearly polarized and the medium is isotropic, $\mathbf{j}_\mathrm{c}$ is
parallel to the $x$--axis. The current density $j_x(t)$ and the charge
transferred along the laser polarization are shown in
Fig.~\ref{fieldonlyfig}(b) together with the electric field of the
laser pulse, which is depicted by the dashed black line.

In Fig.~\ref{fieldonlyfig2}, we show the effects of electron--phonon
scattering on the dynamics of a conduction-band wave packet initially
centred at $\mathbf{k}=0$. A comparison with Fig.~\ref{fieldonlyfig},
where scattering was neglected, reveals that the electron--phonon
interaction disperses the electron wave packet, but it has a
relatively weak effect on $j_x(t)$, decreasing the amplitude of its
oscillations by merely 20\%.
\begin{figure}[htb]
\begin{center}
\includegraphics[width=0.45\textwidth]{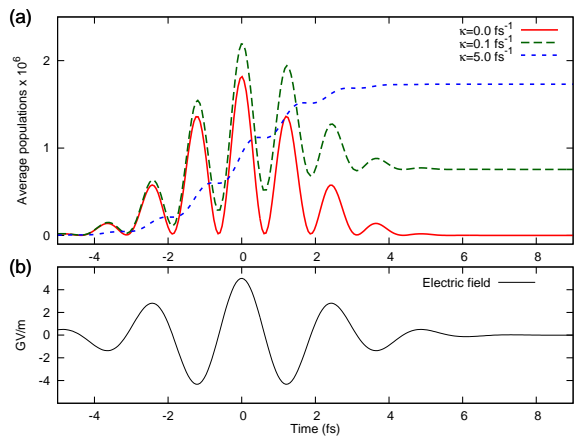}
\caption{Average conduction-band population \eqref{nc} for different
  interband coherence relaxation rates $\kappa$. For $\kappa=5.0\
  \mbox{fs}^{-1}$ (dashed blue curve), we plot $0.05 \langle
  n_\mathrm{c}\rangle$. The parameters are $\mathcal{E}_0=5\
  \mbox{GV}\,\mbox{m}^{-1}$, $\varphi_{\mathrm{CEP}}=0, \tau=2.3\
  \mbox{fs}$, $\gamma=0.3\ \mbox{fs}^{-1}$.}
\label{excfig}
\end{center}
\end{figure}

\subsection{Nonresonant interband excitations}
In order to expose a medium to a very intense laser field without
destroying it, the medium must be possibly transparent, so that little
energy will remain in the medium after the interaction with the pulse.
This implies that the laser frequency $\omega_0$ should be much
smaller than the band gap. This is why we are interested in studying
nonresonant excitations. With our laser parameters, it takes more than
five laser photons to bridge the band gap of $\mbox{SiO}_2$.

The dynamics of interband excitations predicted by our model are shown
in Fig.~\ref{excfig}. For $\kappa=0$, Eq.~\eqref{bloch}
describes a completely coherent excitation process, with the average
population in the conduction band being roughly proportional to the
laser intensity. These periodic excitations and deexcitations are
sometimes referred to as 'virtual excitations', as they largely
represent distortions of initial electronic states. Both virtual and
real excitations contribute to photocurrents~\cite{Yablonovitch1989}.
In the opposite extreme, if we assume an unrealistically fast
decoherence $\kappa=5\ \mbox{fs}^{-1}$, the conduction band population
becomes a monotonically increasing step-like function of time. Little
is known about ultrafast dephasing in $\mbox{SiO}_2$, so we use
$\kappa=0.1\ \mbox{fs}^{-1}$ for our further simulations as a value
that corresponds to an intermediate regime of interband excitations.

\subsection{Laser-driven motion of photoexcited charge carriers
and the effect of the carrier--envelope phase}
Here, we investigate the outcomes of our model with all three terms of
Eq.~\eqref{formaleq} being taken into account. The most interesting result
of combining excitation and laser-driven motion is that this may
result in a nonzero charge $Q$ transferred by a laser pulse---a
transport effect that does not occur in simulations neglecting
interband transitions, as Figs.~\ref{fieldonlyfig} and \ref{fieldonlyfig2}
illustrate. In this section, we show that this is
a combination of interband transitions and a dispersion law that
determines the transferred charge.

\begin{figure}[htb]
\begin{center}
\includegraphics[width=0.45\textwidth]{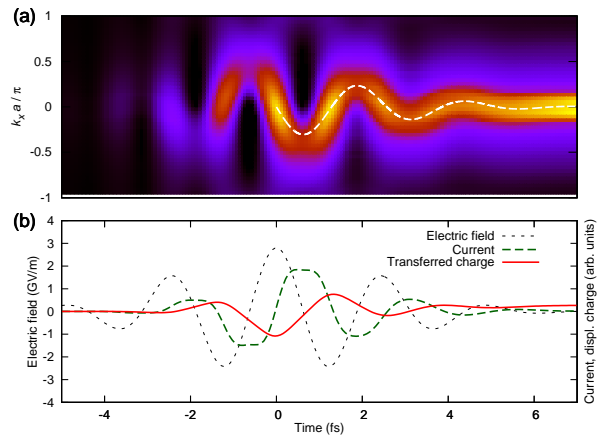}
\caption{(a) Time evolution of the conduction-band electron population
  that emerges due to multiphoton excitations. The cross section in
  the $k_x$ direction (parallel to the polarization of the laser field) is
  shown for the following parameters: $\mathcal{E}_0=3\
  \mbox{GV}\,\mbox{m}^{-1}, \varphi_{\mathrm{CEP}}=0, \tau=2.3\
  \mbox{fs}, \kappa=0.1\ \mbox{fs}^{-1}$, $\gamma=0.3\
  \mbox{fs}^{-1}$. The dashed white line represents a semiclassical
  trajectory released at $t=0$ with $\mathbf{k}=0$. (b) The electric field
  of the laser pulse $\mathcal{E}_x(t)$, the induced current density
  $j_x(t)$ and the transferred charge $Q(t)$ as functions of
  time. \label{lowcosfig1}}
\end{center}
\end{figure}

\begin{figure}[htb]
  \begin{center}
    \includegraphics[width=0.45\textwidth]{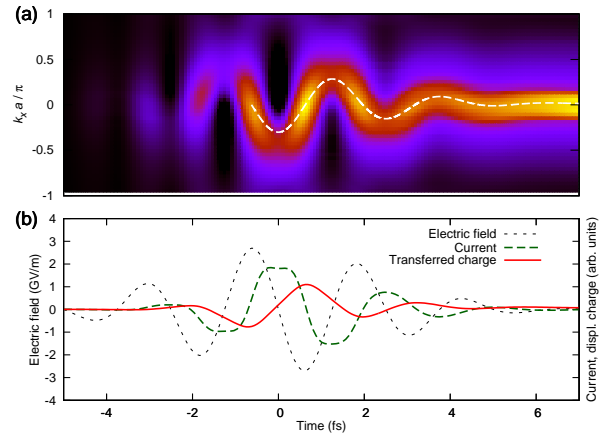}
    \caption{The same as Fig.~\ref{lowcosfig1}, but for a sine pulse
      ($\varphi_{\mathrm{CEP}}=\pi/2$). The semiclassical trajectory
      begins at the first main peak of the laser field,
      i.e.\ at $t=-\pi (2\omega_0)^{-1} \approx -0.6\ \mbox{fs}$, and
      it trespasses $\mathbf{k}=0$ at the peak of the next half-cycle,
      which is the starting point of the second dominant electron
      trajectory.
      \label{lowsinfig}}
  \end{center}
\end{figure}

In the upper panels of Figs.~\ref{lowcosfig1} and \ref{lowsinfig},
we show $n_\mathrm{c}(k_x,t)$ obtained in simulations that account for
all the relevant processes: multiphoton excitations, light-driven
motion of charge carriers, dephasing and electron--phonon scattering.
The simulation were performed for a 'cosine'
($\varphi_\mathrm{CEP}=0$) and a 'sine' ($\varphi_\mathrm{CEP}=\pi/2$)
laser pulses, respectively. The lower panels of the figures show the
electric field of the laser pulse $\mathcal{E}_x(t)$, the induced
current density $j_x(t)$ and the transferred charge $Q(t)$ as
functions of time. Because of scattering, the current density at the
end of each simulation quickly approaches zero, but the final value of
the transferred charge is in general nonzero, and it has a
significantly higher value for the cosine pulse
(Fig.~\ref{lowcosfig1}).

The dashed lines in Figs.~\ref{lowcosfig1}(a) and \ref{lowsinfig}(a)
represent 'semiclassical electron trajectories' released at a peak of
the electric field, which are solutions of Eq.~\eqref{acceleration_theorem}
with the initial condition $\mathbf{k}(t_0)=0$, the initial time being
$t_0=0$ for the cosine pulse and $t_0=-\pi (2 \omega_0)^{-1}$ for the
sine pulse. Even without taking scattering into account, this
semiclassical analysis can be used to explain many features observed
in our calculations. In this picture, the contribution from a
particular electron to the transferred charge at a final time
$t_\mathrm{max}$ is determined by the semiclassical electron
displacement:
\begin{equation}
  \mathbf{s}(t_0) = \int_{t_0}^{t_\mathrm{max}} \mathbf{v}_\mathrm{c}\bigl(\mathbf{k}(t)\bigr)\,\rmd t,
  \label{displacement}
\end{equation}
where $t_0$ is a time when the electron appeared in the conduction
band, $\mathbf{k}(t)$ satisfies Eq.~\eqref{acceleration_theorem},
$\mathbf{v}_\mathrm{c}(\mathbf{k})=\hbar^{-1}\nabla_{\mathbf{k}}
\energy_\mathrm{c}(\mathbf{k})$ and an implicit assumption was made that
the initial velocity of the electron is zero. In the case of a short
cosine pulse, the central half-cycle of the electric field has a
significantly higher amplitude than any other half-cycle.
Consequently, there is one dominant semiclassical trajectory that
starts at the peak of the main half-cycle. A simple calculation shows
that, in our example, the semiclassical displacement associated with
this trajectory is negative, which agrees with the positive final
transferred charge in Fig.~\ref{lowcosfig1}. For a short sine pulse,
there are two dominant trajectories which start at the peaks of the
two most pronounced half-cycles of the electric field. The electron
displacements associated with these trajectories have opposite signs
and add 'destructively' in the case shown in Fig.~\ref{lowsinfig}.
This explains why the magnitude of the net transferred charge is
considerably smaller for $\varphi_{\mathrm{CEP}}=\pi/2$ than for
$\varphi_{\mathrm{CEP}}=0$.

\begin{figure}[htb]
\begin{center}
\includegraphics[width=0.45\textwidth]{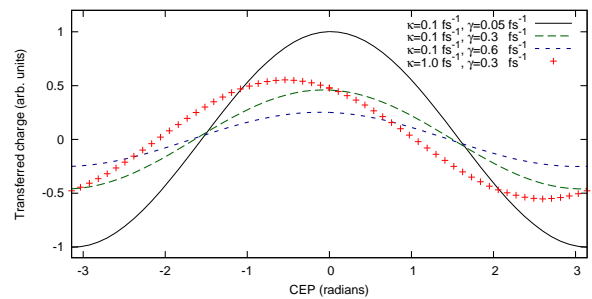}
\caption{The dependence of the total charge transferred by a laser
  pulse on its carrier--envelope phase (CEP). The parameters are
  $\mathcal{E}_0=2\ \mbox{GV}\,\mbox{m}^{-1}$, $\tau=2.3\ \mbox{fs}$,
  $\gamma=0.3\ \mbox{fs}^{-1}$. The curve corresponding to
  $\kappa=0.1\ \mbox{fs}^{-1},\ \gamma=0.05\ \mbox{fs}^{-1}$ has been normalized; the
  other ones have been scaled by the same number, i.e.\ the amplitudes
  of the curves relative to each other are kept fixed. Note that we
  use $\kappa=0.1\ \mbox{fs}^{-1},\ \gamma=0.3\ \mbox{fs}^{-1}$ in the following.
  \label{lowcepfig}}
\end{center}
\end{figure}

We further elaborate on the role of the CEP in Fig.~\ref{lowcepfig},
where we plot $Q(\varphi_{\mathrm{CEP}})$ for different values of
$\kappa$ and $\gamma$. From this figure, one can see that
$Q(\varphi_{\mathrm{CEP}}+\pi)=-Q(\varphi_{\mathrm{CEP}})$, which is a
direct consequence of symmetry: adding $\pi$ to the CEP is equivalent
to substituting $\mathcal{E}(t)$ with $-\mathcal{E}(t)$, which is
equivalent to replacing $x$ with $-x$ in our symmetric arrangement.
One also infers from Fig.~\ref{lowcepfig} that the relaxation rate
of interband coherences influences the positions of minima and maxima
of $Q(\varphi_{\mathrm{CEP}})$, which is not the case for the phonon
scattering rate $\gamma$.

\begin{figure}[htb]
\begin{center}
\includegraphics[width=0.45\textwidth]{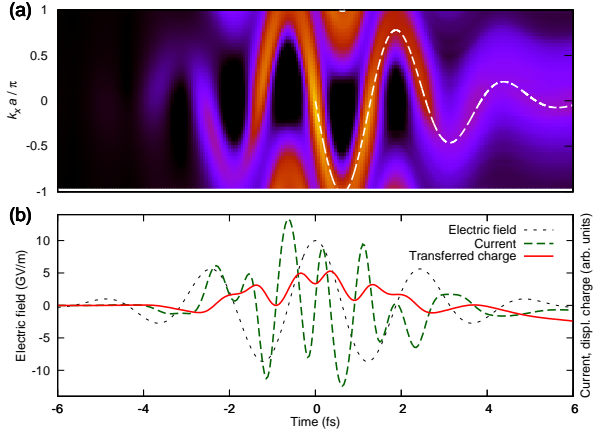}
\caption{The effect of dynamical Bloch oscillations in a simulation
  with the following parameters: $\mathcal{E}_0=10\
  \mbox{GV}\,\mbox{m}^{-1}$, $\varphi_{\mathrm{CEP}}=0$, $\tau=2.3\
  \mbox{fs}$, $\kappa=0.1\ \mbox{fs}^{-1}$, $\gamma=0.3\
  \mbox{fs}^{-1}$. (a) Time evolution of the conduction-band electron
  population $n_\mathrm{c}(k_x,t)$ for a laser field with an amplitude
  sufficient for accelerating excited electrons to the edges of the
  Brillouin zone. The white dashed line represent the solution of
  Eq.~\eqref{acceleration_theorem} with $\mathbf{k}(0)=0$. (b) The electric
  field of the laser pulse $\mathcal{E}_x(t)$, the induced current
  density $j_x$ and the transferred charge $Q(t)$.}
\label{largecosfig}
\end{center}
\end{figure}
Increasing the amplitude of the input laser fields, we reach the
regime where even electrons initially excited in the middle of the
Brillouin zone experience reflections at its edges during the laser
pulse---dynamical Bloch oscillations (DBOs) take place. This is
illustrated in Fig.~\ref{largecosfig}, where we plot the outcomes of
a simulation with a laser pulse that has a peak amplitude of the
electric field of 10 $\mbox{GV}\,\mbox{m}^{-1}$, the other parameters
being the same as in the previous simulations. DBOs occur in spite of
the fact that we use a fairly large value for the the electron--phonon
scattering rate ($\gamma=0.3\ \mbox{fs}^{-1}$). In contrast to the
case of a low field (Fig.~\ref{lowcosfig1}), the final transferred
charge is now negative, which can be interpreted in terms of the
semiclassical electron displacement. The dominant electron trajectory,
shown as a dashed white line, crosses the lower edge of the Brillouin
zone, but it does not reach its upper edge. Since
$\mathbf{v}_\mathrm{c}(\mathbf{k})=0$ at the edges of the Brillouin zone, the
electron is displaced maximally during the time when $k_x>0$.
Consequently, the final electron displacement is positive, and the
transferred charge is negative.

Note that the temporal evolution of the current in
Fig.~\ref{largecosfig} contains frequency components higher than
those of the driving field. The same effect in a different parameter
regime was reported to contribute to the generation of nonperturbative
high-order harmonics in solids~\cite{Ghimire_NP_2011}.

\begin{figure}[htb]
\begin{center}
\includegraphics[width=0.45\textwidth]{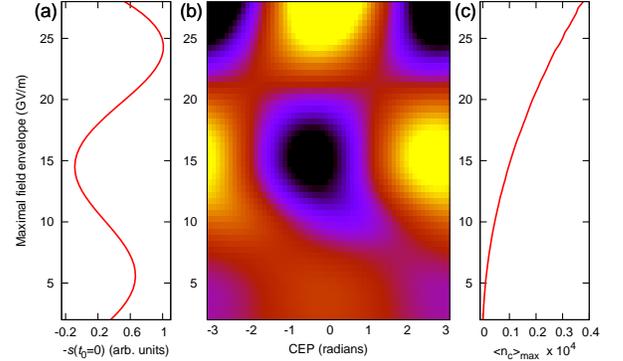}\\[-5mm]
\caption{(a) The semiclassical electron displacement
  (\ref{displacement}) times electron charge for $\varphi_\mathrm{CEP}=0$, $t_0=0$ and
  $t_\mathrm{max}=15\ \mbox{fs}$. The parameters of the laser pulse
  are $\tau=2.3\ \mbox{fs}$ and $\omega_0=2.51\ \mbox{fs}^{-1}$. (b)
  The normalized transferred charge $\tilde{Q}$, defined by
  Eq.~\eqref{Qtilde}, as a function of the CEP and the amplitude of the
  laser pulse. For dephasing and relaxation, we used $\kappa=0.1\
  \mbox{fs}^{-1}$ and $\gamma=0.3\ \mbox{fs}^{-1}$. (c) The maximal
  average population in the conduction band $\langle
  n_{\mathrm{c}}\rangle_{\mathrm{max}}(\mathcal{E}_0)$, defined by
  Eq.~\eqref{nmax}, for $\varphi_{\mathrm{CEP}}=0$. }
\label{cepGVfig}
\end{center}
\end{figure}

Figs.~\ref{lowcosfig1} and \ref{largecosfig} demonstrate that DBOs
have a large impact on the CEP dependence of the transferred charge
$Q(\varphi_\mathrm{CEP})$. In Fig.~\ref{cepGVfig}, we investigate it
more systematically by plotting the normalized transferred charge
\begin{equation}
\tilde{Q}(\varphi_{\mathrm{CEP}},\mathcal{E}_0) =
\frac{Q(\varphi_{\mathrm{CEP}},\mathcal{E}_0)}{\langle n_{\mathrm{c}}\rangle_{\mathrm{max}}(\mathcal{E}_0)}
\label{Qtilde}
\end{equation}
for different laser intensities. The $\mathcal{E}_0$-dependent normalization factor
\begin{equation}
\langle n_{\mathrm{c}}\rangle_{\mathrm{max}}(\mathcal{E}_0)=\Omega^{-1} \max_{t} \int_{\mathrm{BZ}} n_\mathrm{c}(\mathbf{k},\mathcal{E}_0, t, \varphi_{\mathrm{CEP}}=0)\,\rmd^2 \mathbf{k},
\label{nmax}
\end{equation}
which is plotted in Fig.~\ref{cepGVfig}(c), denotes the maximum of
the time dependent average population in the conduction band for
$\varphi_{\mathrm{CEP}}=0$.

One immediately observes that the extrema of
$Q(\varphi_{\mathrm{CEP}})$ shift as the laser intensity increases. By
inspecting these simulations, we identified DBOs for peak laser fields
above 7~$\mbox{GV}\,\mbox{m}^{-1}$. Below this limit,
$Q(\varphi_{\mathrm{CEP}})$ is approximately proportional to
$\cos(\varphi_{\mathrm{CEP}})$, and only the amplitude of
$Q(\varphi_{\mathrm{CEP}})$ increases with the peak laser intensity.
For intensities high enough to induce DBOs, the extrema of
$Q(\varphi_{\mathrm{CEP}})$ change their positions. For
$\mathcal{E}_0=15\ \mbox{GV}\,\mbox{m}^{-1}$, maxima and minima
exchange their places.

To relate these observations to the semiclassical analysis, we focus
on the cosine pulse and plot the electron displacement at
$t_\mathrm{max}=15\ \mbox{fs}$ as a function of the peak electric
field in Fig.~\ref{cepGVfig}(a). We see that this analysis
qualitatively explains our numerical results:
$Q(\varphi_\mathrm{CEP}=0,\mathcal{E}_0)$ and $-s(t_0=0,\mathcal{E}_0)$
have their extrema at approximately the same values of the peak
electric field $\mathcal{E}_0$, the minus sign being due to the
negative electron charge.

Finally, let us return to our model given by Eq.~\eqref{formaleq}
and summarize the physical role of the various processes. Multiphoton
excitation and laser-driven motion in the conduction and valence bands
are essential for the CEP dependence of the transferred charge, as
well as for the appearance of dynamical Bloch oscillations. LO phonon
scattering, however, is a process that competes with laser-driven
interband motion, and could possibly render the detection of DBOs
impossible. According to our calculation, this is not the case even
for the large scattering rates known for $\mbox{SiO}_2$~\cite{Hughes1973,Fischetti1985}.

One of our major approximations is using just two bands: a valence and
a conduction one. This approximation was made not because we can
exclude the involvement of higher conduction or lower valence bands,
but because our intention was to clarify the role of Bloch
oscillations. In a more realistic description, crossing the edge of a
Brillouin zone does not necessarily imply Bragg-like scattering of an
electron---transitions to other bands lead to more complicated
dynamics~\cite{Breid2006}, which may, for example, smear the
dependencies shown in Fig.~\ref{cepGVfig}, especially in the region
of the highest intensities. To which extent this happens in a
particular measurement will depend on the chosen material, sample
preparation, laser wavelength, and other parameters. Even though we
neglect these transitions, our results may assist the interpretation
of future measurements by recognizing certain features as evidence of
Bloch oscillations.

\section{Summary}
Using a phenomenological model, we have investigated multiphoton
injection and laser-driven motion of charge carriers in a wide-gap
bulk dielectric exposed to intense few-cycle laser pulses. In
comparison with more rigorous quantum models~\cite{Schiffrin2013,Apalkov2012,Kruchinin2013}, where
different physical phenomena are relatively difficult to disentangle,
our approach lends itself to clarifying the roles played by various
processes. Our most important finding is that whenever a laser field
drives electrons close to or beyond the edges of the Brillouin zone,
Bragg-like reflections of electron waves have a significant
impact on CEP-sensitive measurements like those reported in~Ref.~\onlinecite{Schiffrin2013}
(the phase shift by $\pi$ in
Fig.~\ref{cepGVfig}). At the same time, the role of electron--phonon
scattering is limited to reducing the amount of the transferred charge
without qualitatively affecting CEP dependencies, even if we assume
scattering rates as high as $\gamma \sim 10^{14}\ \mbox{s}^{-1}$.
The fact that electron--phonon
scattering plays a minor role in the high-intensity regime justifies
neglecting it in earlier models~\cite{Schiffrin2013,Kruchinin2013}. Our
final conclusion is that the detection of the total transferred charge
can be used to measure signatures of dynamical Bloch oscillations in
bulk solids. Furthermore, we show that this effect can be
qualitatively explained in terms of the semiclassical electron
displacement evaluated for dominant electron trajectories.

\section*{Acknowledgements}
The authors are indebted to A.~Schiffrin, S.~Kruchinin, T.~Paasch-Colberg, F.~Krausz and,
particularly, N.~Karpowicz for illuminating discussions.
This work was supported by the Hungarian Scientific Research Fund
(OTKA) under Contract No.~T81364 as well as by the
projects T\'{A}MOP-4.2.2.A-11/1/KONV-2012-0060 and
T\'{A}MOP-4.2.2.C-11/1/KONV-2012-0010 supported by the European Union
and co-financed by the European Social Fund. V.~S.~Y. acknowledges
support by the DFG Cluster of Excellence: Munich-Centre for Advanced
Photonics.


\bibliography{Foldi_Bloch_oscillations}

\end{document}